# Nearest-Neighbor Broken Bond Model for Describing the Surface Energies of Transition Metals

Xiao-Yan Li[1,2], Beien Zhu[1,3,*], Yi Gao[1,3,*]

[1]Division of Interfacial Water and Key Laboratory of Interfacial Physics and Technology, Shanghai Institute of Applied Physics, Chinese Academy of Sciences, Shanghai, 201800 China

[2]University of Chinese Academy of Sciences, Beijing, 100049 China

[3]Shanghai Advanced Research Institute, Chinese Academy of Sciences, Shanghai, 201210 China

E-mail: gaoyi@sinap.ac.cn

In our recent works, we used a nearest-neighbor broken bond (NNBB) model to fit the surface energy of Pt, Pd, and Au surfaces [1]. It was verified that this model could describe the surface energies of above noble metals with high accuracy. The applications of this model in real-time simulations of equilibrium nanocrystal transformations are in good agreements with the experimental observations at macroscopic time scales quantitatively. In this work, we want to further verify the generality of the NNBB model for other transition metals. All surface tentions (eV/Å$^2$) and lattice constants (Å) were extracted from the database of Materials Project [2]. The number of missing NN bonds of different slabs from low-index surfaces to open surfaces were counted for both surface atoms and subsurface atoms. As presented below, linear relationships between surface energies and missing bonds number are found for all studied elements. The bond energies ($\varepsilon$) for different metals are obtained through fitting NN bonds and surface energy of different surfaces.

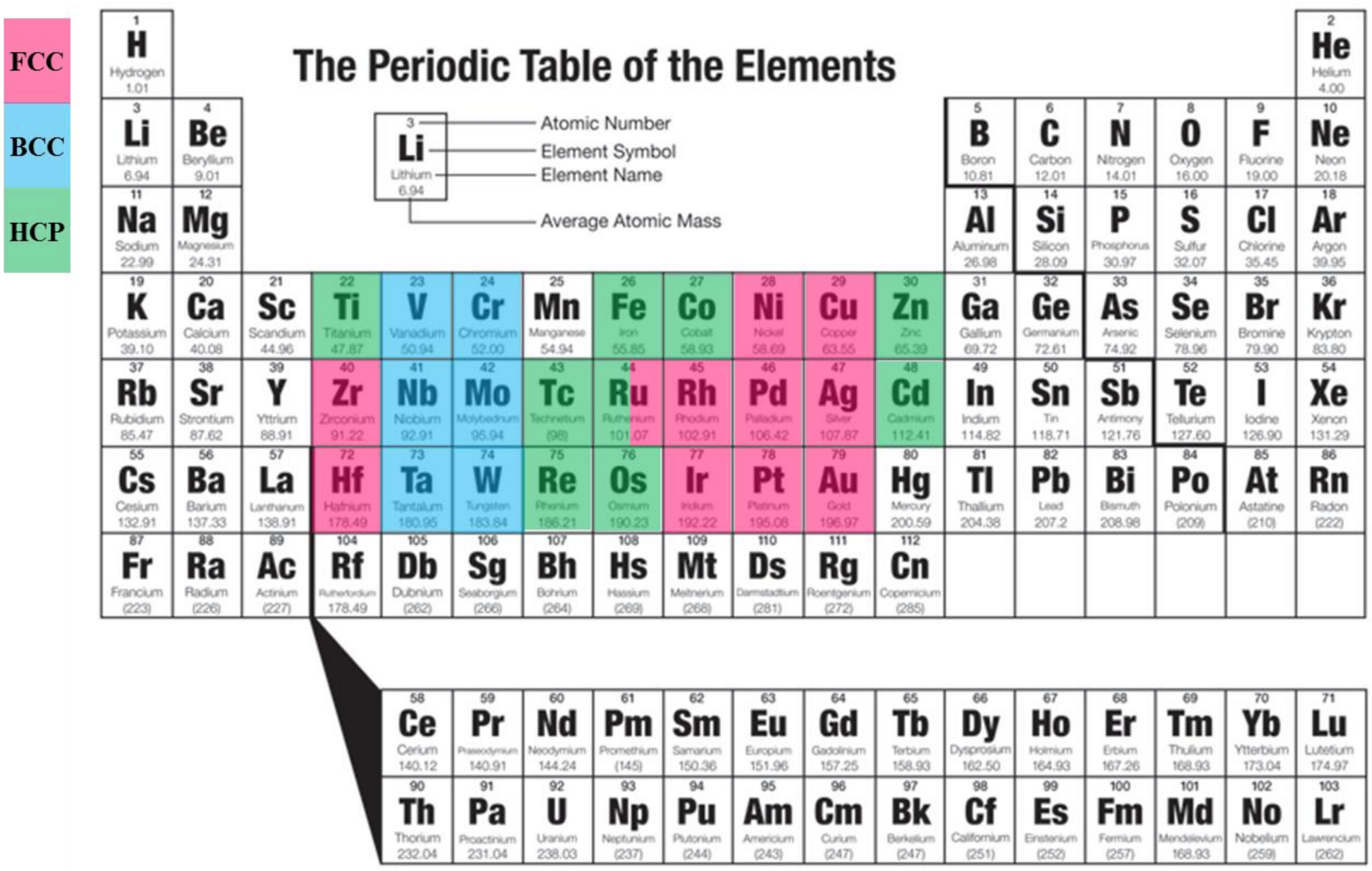

**Figure 1.** The colored elements are considered in this paper. Red, blue and green refers to face centered cubic (FCC), body centered cubic and hexagonal closed-pack structures. The rest elements without color indicate their surface energies are not given in the Materials Project.

Transition metals with different crystal structure are investigated (as shown in Figure 1), namely space group: Fm$\bar{3}$m, Im$\bar{3}$m, P6$_3$/mmc. For FCC and BCC structures, thirteen index surfaces are considered, including (100), (110), (111), (211), (221), (210), (310), (311), (320), (321), (322), (331), and (332). For the hexagonal close-packed (HCP) structures, twelve index surfaces are considered, including (0001), ($10\bar{1}0$),

$(10\bar{1}1)$, $(10\bar{1}2)$, $(11\bar{2}0)$, $(11\bar{2}1)$, $(20\bar{2}1)$, $(21\bar{3}0)$, $(21\bar{3}1)$, $(21\bar{3}2)$, $(2\bar{1}\bar{1}2)$ and $(22\bar{4}1)$. The linear relationships between FCC, BCC, and HCP surface energies and the number of missing NN bonds of different metals are shown in **Figure 2**, **Figure 3** and **Figure 4**, respectively.

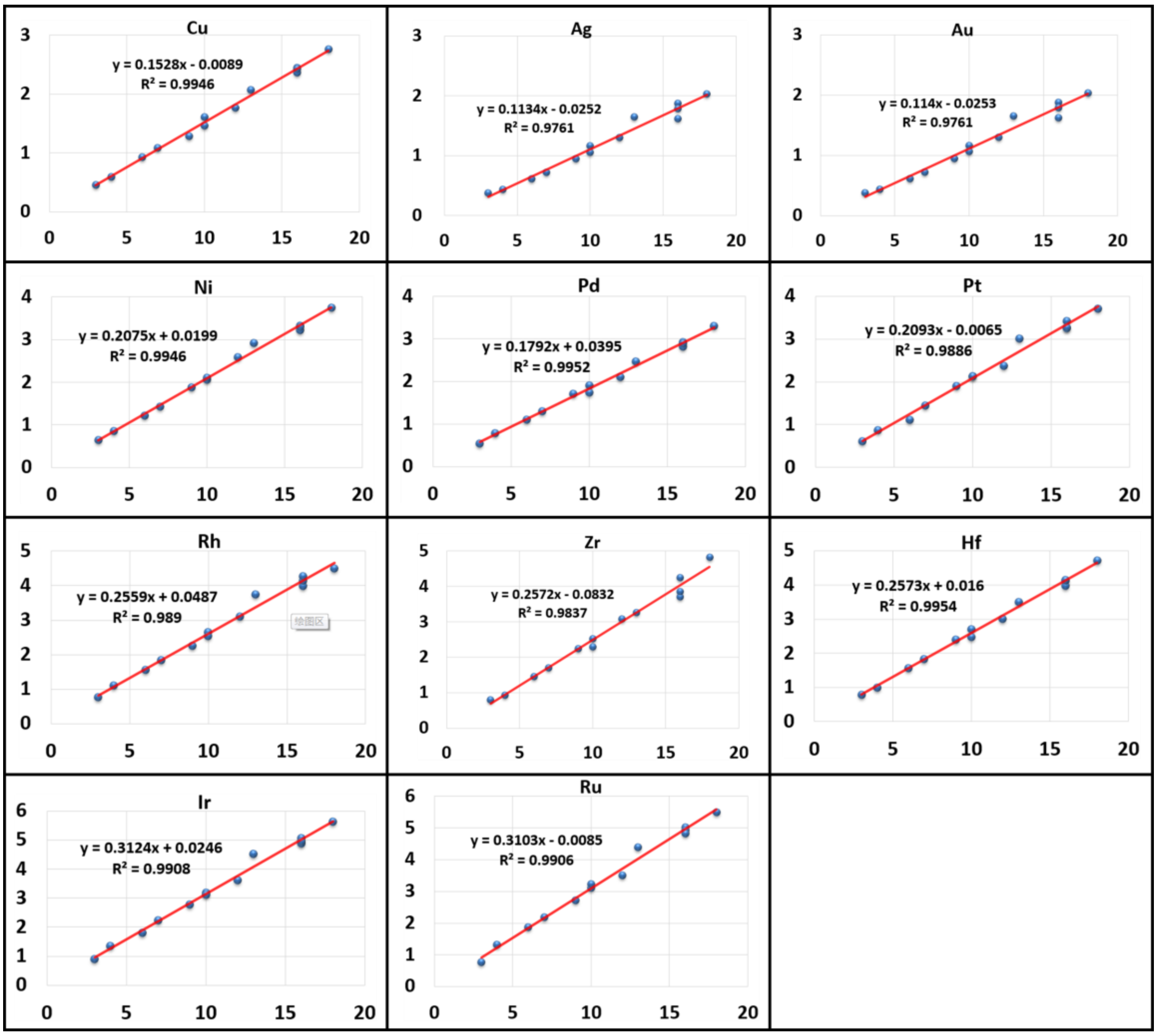

**Figure 2.** The linear relationship between missing NN bonds and surface energies (eV) of FCC structures. The horizontal ordinate refers to the number of missing NN bonds. The vertical ordinate refers to the surface energy. In every subgraph, the corresponding formula is shown using linear fitting.

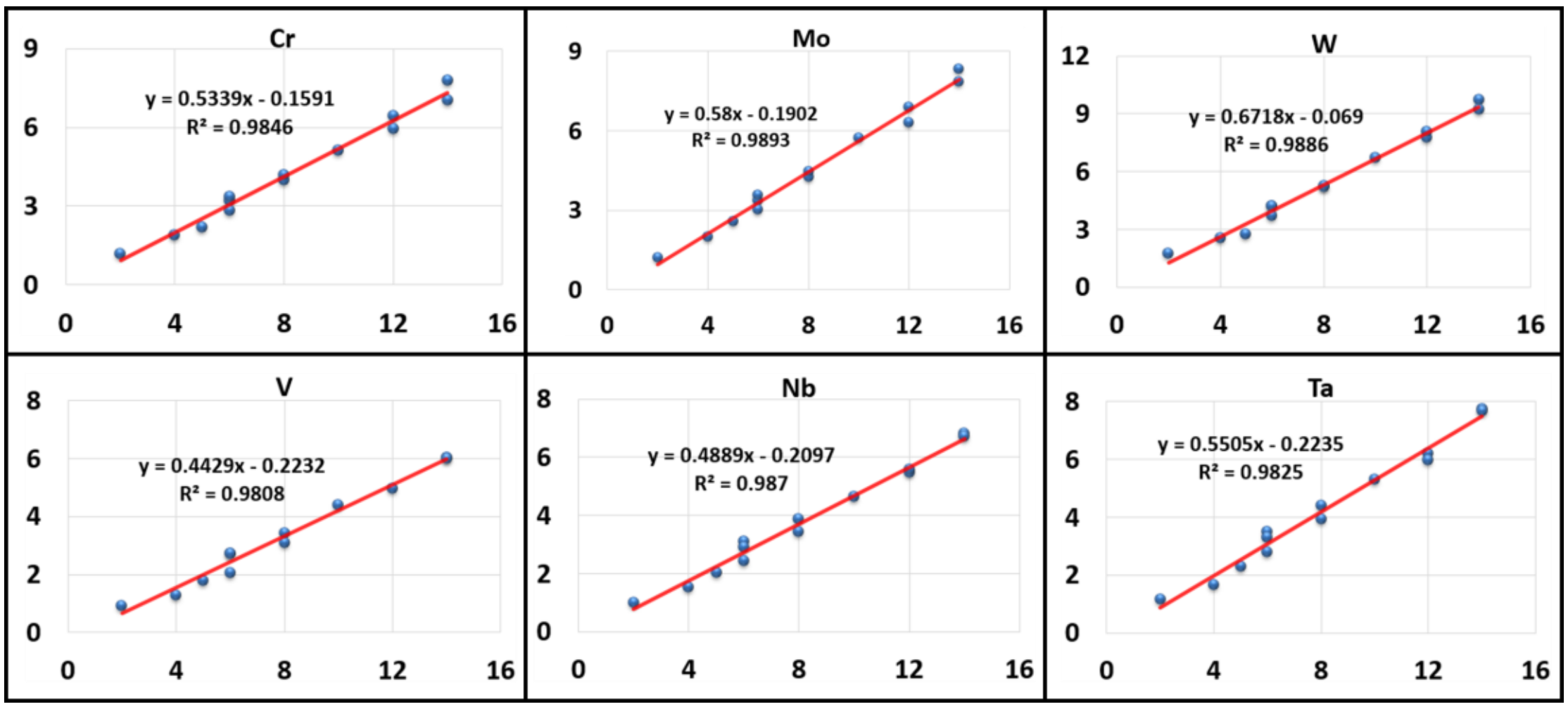

**Figure 3.** The linear relationship between missing NN bonds and surface energies (eV) of BCC structures.

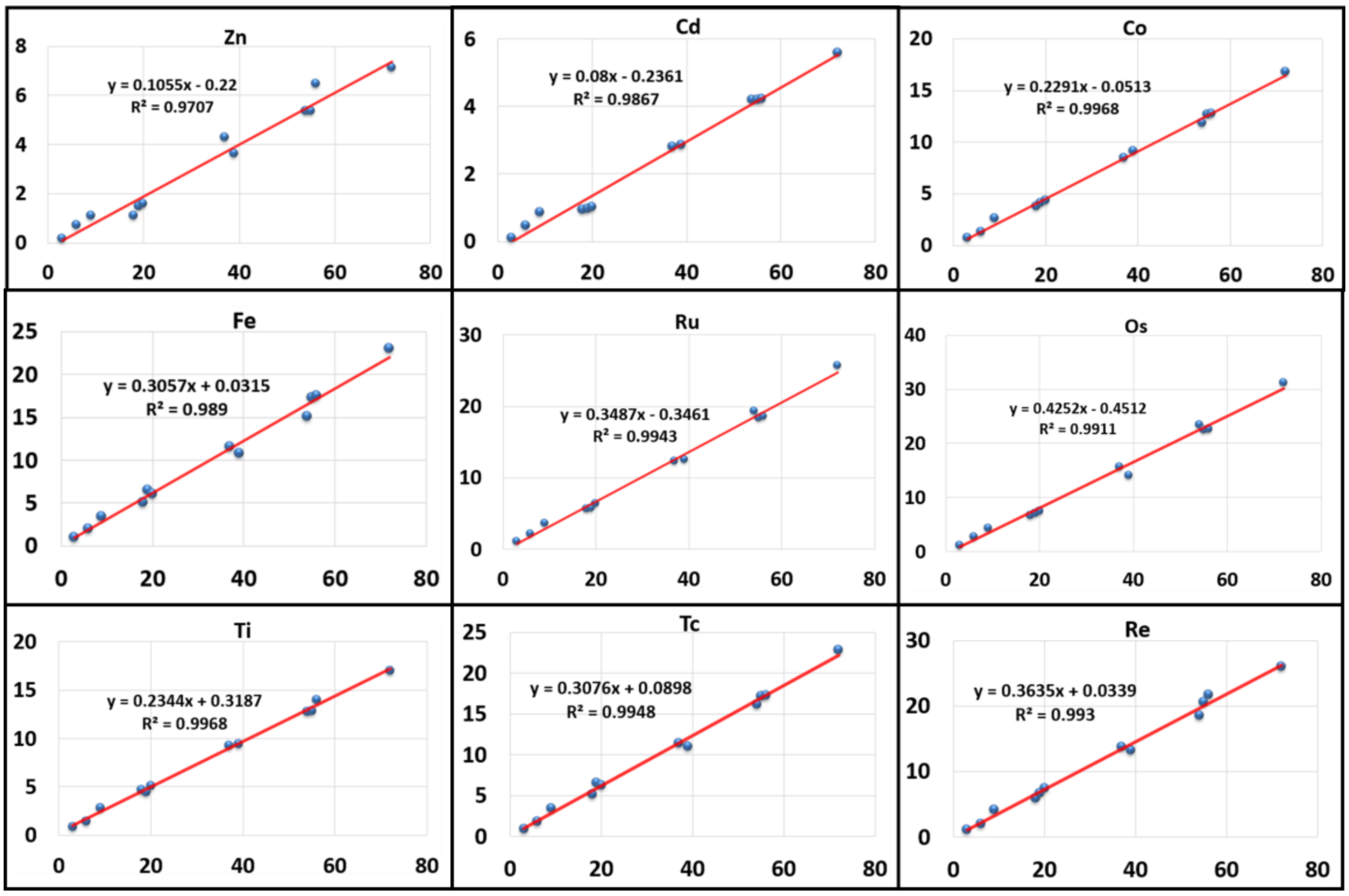

**Figure 4.** The linear relationship between missing NN bonds and surface energies (eV) of HCP structures.

It is assumed that the surface energy of a metal surface is contributed from the missing NN bonds:

$$E_{hkl}^{surf} = -\frac{1}{2} N_{hkl} \varepsilon \qquad (1)$$

$N_{hkl}$ is defined as the number of missing NN bonds when cutting it from bulk and $E_{hkl}^{surf}$ is the surface energy of a unit-cell [*hkl*] surface, which is calculated by multiplying the surface energy per area by the unit surface area. .

Hence, based on fitted linear relationship, the bond energy ($\varepsilon$) of each transition metal with corresponding crystal structure can be obtained, as shown in the **Table 1**, **Table 2** and **Table 3**.

**Table 1.** The fitted bond energy ($\varepsilon$) and corresponding structural lattice constant (FCC) are listed below.

| Element | Lattice constant a (Å) | Bond energy (eV) |
|---|---|---|
| Cu | 3.62 | -0.31 |
| Ag | 4.16 | -0.23 |
| Au | 4.17 | -0.23 |
| Ni | 3.51 | -0.42 |
| Pd | 3.95 | -0.36 |
| Pt | 3.98 | -0.42 |
| Rh | 3.84 | -0.51 |
| Zr | 4.54 | -0.51 |
| Hf | 4.48 | -0.51 |
| Ir | 3.88 | -0.62 |
| Ru | 3.83 | -0.62 |

**Table 2.** The fitted bond energy ( $\varepsilon$ ) and corresponding structural lattice constant (BCC) are listed below.

| Element | Lattice constant a ($\mathring{A}$) | Bond energy (eV) |
|---|---|---|
| Cr | 2.85 | -1.07 |
| Mo | 3.17 | -1.16 |
| W | 3.19 | -1.34 |
| V | 2.93 | -0.59 |
| Nb | 3.32 | -0.97 |
| Ta | 3.32 | -1.10 |

**Table 3.** The fitted bond energy ( $\varepsilon$ ) and corresponding structural lattice constant (HCP) are listed below.

| Element | Lattice constant ($\mathring{A}$) a and c | Bond energy (eV) |
|---|---|---|
| Zn | 2.63 5.21 | -0.21 |
| Cd | 3.01 5.94 | -0.16 |
| Co | 2.50 4.03 | -0.46 |
| Fe | 2.47 3.90 | -0.61 |
| Ru | 2.74 4.29 | -0.70 |
| Os | 2.76 4.36 | -0.85 |
| Ti | 2.93 4.66 | -0.47 |
| Tc | 2.76 4.42 | -0.62 |
| Re | 2.78 4.50 | -0.73 |

The generality of the NN bonding model for the surface energy description is beyond imagination. The physical insight behind this generality is an interesting question to be answered. We would like to open this question to the scientific community for further discussions.